# Braided and Knotted Stocks in the Stock Market: Anticipating the flash crashes

Ovidiu Racorean

Applied Mathematics in Finance Dept., SAV Integrated Systems

Email: rovidiu@sav-integrated.ro

**Abstract**

A simple and elegant arrangement of stock components of a portfolio (market index-DJIA) in a recent paper [1], has led to the construction of **crossing of stocks diagram**. The crossing stocks method revealed hidden remarkable algebraic and geometrical aspects of stock market. The present paper continues to uncover new mathematical structures residing from crossings of stocks diagram by introducing topological properties stock market is endowed with. The crossings of stocks are categorized as overcrossings and undercrossings and interpreted as generators of braid that stocks form in the process of prices quotations in the market. **Topological structure of the stock market** is even richer if the closure of stocks braid is considered, such that it forms a **knot**. To distinguish the kind of knot that stock market forms, Alexander-Conway polynomial and the Jones polynomials are calculated for some knotted stocks. These invariants of knots are important for the future practical applications topological stock market might have. Such application may account of the relation between Jones polynomial and phase transition statistical models to provide a clear way to anticipate the transition of financial markets to the phase that leads to crisis. The resemblance between braided stocks and logic gates of topological quantum computers could quantum encode the stock market behavior.

1. Introduction

The recent paper [1] proposes a way of interpreting the behavior of market indexes as a multivariate time series that results by arranging the index stock components in a simply particular manner.   Prices of the stock index components are arranged in a table in ascending order starting from smaller stock prices, in the left, to companies having higher stock prices, to the right. Coloring every stock prices time series in different a color is the way of keeping track with each stock price quotations in the arranged table. Although simple, this technique of arranging the stocks encodes immense mathematical potential and unlocks some remarkable mathematical objects such as permutations, polytopes, braids and knots, to show up in the financial frame.

The most important concept reside from this particular arrangement of the market index is the **crossing of stocks.** Following the colored stock time series in the arranged table a stocks crossing diagram can be drawn. The ***crossing of stocks diagram*** explicitly shows the moments when the price of a stock comes over or under the price of its neighbor stock in the table.

To exemplify the stocks crossing diagram with examples from the real stock market, in the section 2, the market index is chose to be *Dow Jones Industrial Average* (DJIA) with its components. Prices of a fraction of all 30 components of DJIA are arranged in the manner explained above, starting from CSCO which is the lowest priced stock until the highest, PG, for the market reference date 5/15/2013. To simplify the understanding of these particular diagrams only 4 stocks are retained and will represent the main constituents to exemplify the new mathematical concepts at.

Section 3 introduces the notions of **overcrossing** and **undercrossing** of stocks as generators for the stocks brad formation. Braids are known since the old times but their powerful algebraic properties were first revealed last century in the work of Emil Artin [6]. Intermediated by the generators the crossing of stocks diagram is transformed in braid diagram. Properties of braids as algebraic objects, such as formation of a group remained unchanged in the stock market application.

Another important aspect of stocks braid generators is explored briefly along the section 4. In their time succession generators, noted according to financial needs, form words that could "whisper" some important market information to investors and traders.

An important theorem directly connects braid with **knot,** another remarkable mathematical structure. Accounting to this theorem the braiding of stocks in the market can be represented as **stocks knot**. Knot formation in the stock market is explained in section 5. Once the stocks are knotted an important question arises: How to distinguish between the various knot types the stocks can form in the process of price quotation in the market?

Asking to this question is still an open issue in mathematics, but some remarkable steps forward were made in the work of Alexander, J. Conway, V. Jones and recently by many other researchers. The answer, to stick to our discussion, is to calculate some polynomials that appear when a **skein relation** is applied to the crossings of a knot in order to reduce them to none. Section 6 is dedicated to exemplify the calculation for Alexander-Conway polynomial and Jones polynomial, two of the most known and used knot invariants, for knotted stocks.

Having the stock market all knotted and the polynomial invariant calculated is easy to say what kind of knot is by taking a look in the table of knots. A fragment of the knots table, classified by their Alexander polynomial is shown in annex.

Apart from the beauty of the idea of a ***topological stock market*** the practical applications of this concept must be questioned. I should emphasize here that the present paper is a part of a large research project designed to apply geometry and topology to stock market and many aspects related to practical applications are not yet explored. Still some hints of methods knotting of stocks prices could be of help in financial practice are presented in section 7. The connection of Jones polynomial with statistical mechanics of phase transition models can be exploited to anticipate flash crashes that high frequency trading generates and in a larger extent financial crisis. In such scenario the financial crisis is nothing else than a phase transition from a market having a smooth behavior to a market of a regime prone to sharp, with virtually discontinuous price movements.

The resemblance of the braiding stocks with the logic gates of the topological quantum computer is a second hint about the ways braided and knotted stocks topology could be applied to financial realities. Under this scenario the Jones polynomial of some stocks knot would represent qubits that encode the stock market quantum states in the way quantum computers operate. It might be a simple speculation, but in the down of the cryptographic money era, opened by bitcon, it could have some important connotations.

Leaving aside these speculations, a lot of work is still to be done to uncover the remarkable mathematical objects hidden behind the simple crossing of stocks diagram.

2. **Crossing of stocks diagram**

One of the most important finding in [1] is the prescription to arrange the stock components of a market index (Dow Jones Industrial Average – DJIA) in a manner that allow the presence of relations between stocks to show up as **crossing of stocks**. It is suffice to tell that could represent a way to analyze the **multivariate time series**. Although simple, this technique of arranging the stocks encodes immense mathematical potential and highlights some remarkable mathematical objects such as permutations, polytopes, braids and knots, to show up in the financial frame.

Although the crossings of stocks are of crucial importance, the paper [1] has not devoted to the subject the attention that it deserves. This section constitutes an attempt to straighten this "injustice" by presenting in more details the *crossings of stocks diagram* and its powerful mathematical advantages.

The index of interest is choosing to be the *Dow Jones Industrial Average* (DJIA). A fraction of price quotations for some DJIA components are shown in table 1 as daily closing prices for a period in 2013 between 5/15/2013 and 6/7/2013.

| Date | CSCO | GE | INTC | PFE | MSFT | T | KO | MRK | VZ | JPM | DD | DIS | NKE | UNH | AXP | HD | WMT | PG |
|---|---|---|---|---|---|---|---|---|---|---|---|---|---|---|---|---|---|---|
| 6/7/2013 | 24.5 | 23.9 | 24.6 | 28.3 | 35.7 | 35.5 | 41.4 | 48.2 | 50.2 | 54.3 | 55.4 | 64.9 | 62.8 | 62.6 | 78 | 78.7 | 76.3 | 77.8 |
| 6/6/2013 | 24.6 | 23.4 | 24.7 | 28.1 | 35 | 35.8 | 40.3 | 48.6 | 50 | 53.5 | 54.8 | 63.1 | 62.2 | 61.9 | 76.2 | 77.3 | 75.6 | 76.8 |
| 6/5/2013 | 24.3 | 23.3 | 24.7 | 27.5 | 34.8 | 35.3 | 40.7 | 48.7 | 48.3 | 53 | 54.6 | 63.1 | 61.8 | 61.8 | 74.8 | 75.1 | 75.3 | 76.7 |
| 6/4/2013 | 24.4 | 23.7 | 25.4 | 27.7 | 35 | 35.7 | 41.4 | 49.4 | 48.8 | 54 | 55.8 | 64.4 | 62.8 | 62.4 | 76.1 | 76.6 | 75.9 | 77.4 |
| 6/3/2013 | 24.4 | 23.6 | 25.2 | 27.8 | 35.6 | 35.1 | 40.3 | 48.5 | 48.7 | 54.5 | 56.1 | 63.8 | 63 | 62.8 | 76.5 | 79.1 | 75.7 | 77.7 |
| 5/31/2013 | 24.1 | 23.3 | 24.3 | 27.2 | 34.9 | 35 | 40 | 46.7 | 48.5 | 54.6 | 55.8 | 63.1 | 61.7 | 62.6 | 75.7 | 78.7 | 74.8 | 76.8 |
| 5/30/2013 | 24.4 | 23.6 | 24.2 | 28.3 | 35 | 35.5 | 40.3 | 47.1 | 49.1 | 55.6 | 56.3 | 64.7 | 62.4 | 64.7 | 76.1 | 79.4 | 75.6 | 79.1 |
| 5/29/2013 | 24.1 | 23.6 | 24.3 | 28.3 | 34.9 | 35.9 | 41.4 | 46.9 | 49.6 | 54.7 | 56 | 66.3 | 62.9 | 63.4 | 75.8 | 79.5 | 76.2 | 78.9 |
| 5/28/2013 | 23.9 | 23.6 | 24.1 | 29 | 35 | 36.2 | 42.6 | 47.6 | 50.3 | 54.6 | 55.9 | 66.7 | 63.3 | 63.3 | 76.2 | 79.8 | 77.3 | 80.9 |
| 5/24/2013 | 23.5 | 23.5 | 23.9 | 29 | 34.3 | 36.8 | 42.2 | 47.2 | 51.4 | 53.7 | 55.4 | 65.5 | 62.8 | 62.1 | 75.3 | 79 | 77.3 | 81.9 |
| 5/23/2013 | 23.5 | 23.7 | 24.1 | 29.1 | 34.2 | 36.7 | 41.9 | 47.3 | 51.9 | 53.4 | 55.4 | 65.2 | 63.3 | 62.4 | 74.7 | 78.9 | 76.3 | 78.7 |
| 5/22/2013 | 23.3 | 23.9 | 24.1 | 29.3 | 34.6 | 36.6 | 42.3 | 46.7 | 51.5 | 53.6 | 55.6 | 65.6 | 64.5 | 62.3 | 74.4 | 79.7 | 77 | 78.8 |
| 5/21/2013 | 24 | 23.7 | 24.2 | 28.8 | 34.9 | 36.9 | 42.3 | 47.3 | 52.1 | 53 | 56.4 | 65.8 | 65.2 | 62.9 | 75.1 | 78.7 | 77.4 | 78.8 |
| 5/20/2013 | 24 | 23.6 | 24.1 | 28.7 | 35.1 | 37.2 | 42.4 | 45.2 | 52.7 | 52.3 | 55.9 | 66.1 | 65.3 | 62.6 | 74.4 | 76.8 | 77.4 | 79.1 |
| 5/17/2013 | 24.2 | 23.5 | 24 | 29 | 34.9 | 37.4 | 43 | 46 | 53.4 | 52.3 | 55.9 | 66.6 | 65.3 | 62.8 | 73.3 | 76.9 | 77.9 | 80 |
| 5/16/2013 | 23.9 | 23.3 | 23.9 | 29.3 | 34.1 | 37.4 | 43.1 | 46.4 | 53.2 | 51 | 55.5 | 66.5 | 64.4 | 62.1 | 72.2 | 76.8 | 78.5 | 80.2 |
| 5/15/2013 | 21.2 | 23.2 | 24.2 | 29.6 | 33.9 | 37.5 | 42.9 | 46.7 | 53.6 | 51.1 | 55.6 | 67.7 | 65.8 | 61.6 | 72.8 | 77.9 | 79.9 | 80.7 |

**Table 1**. A fraction of the DJIA index components sorted by price quotations from the left to the right at 05/15/2013.

It can be easily seen that the DJIA stock components in the table 1 are arranged in ascending order from the stock with the smallest price quotation (CSCO) at the left to the stock with the highest price (PG) at the right at the start date 5/15/2013.This will be a rule of arranging the stock components of the DJIA index.

Notice that for others rows of the table this rule is not applied, such that next day, at 5/16/2013, for example, the price of CSCO came over the price of GE.

To simplify the exposition only four stock components of DJIA are retain further, AXP, HD, WMT and PG.  The number of stocks is chosen such that the discussion should be neither trivial, nor too complex.

The price quotations for the chosen four DJIA components are arranged in ascending order from the stock with the smallest price (AXP) on the left to the stock having the highest price (PG) on the right at the starting date 5/15/2013. The time series of prices for every stock is colored in a different color as is shown in figure 1 a). The arrangement of stocks from the left to the right in ascending order of prices is preserved for every row in the table, say for every trading day. In this manner the stocks prices will be shifted from their initial positions (see figure 1 a), at the right or left, every time the price of one of the four stocks comes under or over the price of the neighbor stock, put it in other words every time the *stocks are crossing*. Figure 1 b) depicted the sorted prices of stocks in ascending order from the left to the right, and the crossings of stocks become very clear by following the stocks colors.

| Date | AXP | HD | WMT | PG | | Date | AXP | HD | WMT | PG |
|---|---|---|---|---|---|---|---|---|---|---|
| 6/7/2013 | 78.04 | 78.74 | 76.33 | 77.75 | | 6/7/2013 | 76.33 | 77.75 | 78.04 | 78.74 |
| 6/6/2013 | 76.24 | 77.26 | 75.63 | 76.82 | | 6/6/2013 | 75.63 | 76.24 | 76.82 | 77.26 |
| 6/5/2013 | 74.76 | 75.1 | 75.25 | 76.66 | | 6/5/2013 | 74.76 | 75.25 | 75.1 | 76.66 |
| 6/4/2013 | 76.06 | 76.63 | 75.94 | 77.37 | | 6/4/2013 | 75.94 | 76.06 | 76.63 | 77.37 |
| 6/3/2013 | 76.47 | 79.08 | 75.69 | 77.66 | | 6/3/2013 | 75.69 | 76.47 | 77.66 | 79.08 |
| 5/31/2013 | 75.71 | 78.66 | 74.84 | 76.76 | | 5/31/2013 | 74.84 | 75.71 | 76.76 | 78.66 |
| 5/30/2013 | 76.14 | 79.44 | 75.63 | 79.09 | | 5/30/2013 | 75.63 | 76.14 | 79.09 | 79.44 |
| 5/29/2013 | 75.83 | 79.49 | 76.23 | 78.9 | | 5/29/2013 | 75.83 | 76.23 | 78.9 | 79.49 |
| 5/28/2013 | 76.16 | 79.82 | 77.32 | 80.86 | | 5/28/2013 | 76.16 | 77.32 | 79.82 | 80.86 |
| 5/24/2013 | 75.27 | 78.99 | 77.31 | 81.88 | | 5/24/2013 | 75.27 | 77.31 | 78.99 | 81.88 |
| 5/23/2013 | 74.69 | 78.91 | 76.33 | 78.7 | | 5/23/2013 | 74.69 | 76.33 | 78.7 | 78.91 |
| 5/22/2013 | 74.44 | 79.69 | 77.03 | 78.82 | | 5/22/2013 | 74.44 | 77.03 | 78.82 | 79.69 |
| 5/21/2013 | 75.11 | 78.71 | 77.39 | 78.8 | | 5/21/2013 | 75.11 | 77.39 | 78.71 | 78.8 |
| 5/20/2013 | 74.4 | 76.76 | 77.4 | 79.09 | | 5/20/2013 | 74.4 | 76.76 | 77.4 | 79.09 |
| 5/17/2013 | 73.32 | 76.86 | 77.87 | 80.02 | | 5/17/2013 | 73.32 | 76.86 | 77.87 | 80.02 |
| 5/16/2013 | 72.23 | 76.75 | 78.5 | 80.2 | | 5/16/2013 | 72.23 | 76.75 | 78.5 | 80.2 |
| 5/15/2013 | 72.78 | 77.88 | 79.86 | 80.68 | | 5/15/2013 | 72.78 | 77.88 | 79.86 | 80.68 |
| | AXP | HD | WMT | PG | | | AXP | HD | WMT | PG |

**Figure 1.**   a) The initial arrangement of stocks       b) The prices of stocks are sorted and show the crossing of stocks

Notice from the figure 1 b) above that from time to time *stocks are crossing.* As an example, it can be seen that at 5/21/2013 the closing prices of HD and WMT are crossing. As it was stated earlier the impact of the crossing on the total value of DJIA is, for now, neglected and the attention is focused only on the colored trajectories the stocks prices take in the DJIA components table.

Bearing in mind only the colored trajectories of stocks prices, totally neglecting the values of quotations, the figure 1 can be transpose in the picture bellow, where the crossings of stocks become very clear:

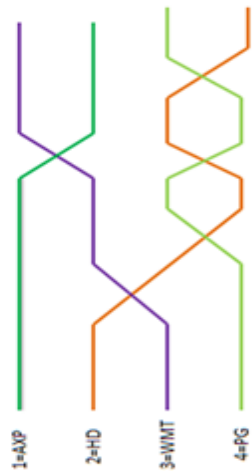

**Figure 2**. Crossings of stocks diagram.

The **crossing of stocks diagram** although simple, hidden remarkable mathematical proprieties which will be explored in the next sections.

3. **The stock market braids**

Braids are known in the day by day experience since the old times. Emil Artin was the first to uncover the powerful mathematical structure of braids in his paper written in the middle of 1920'. He refined the initial mathematical approach of braids in a series of articles until 1947.

Generally, braids consists of $n$ strands having fixed ends that cross which other. To stick to the four stocks example, an arbitrary 4-strands braid is depicted in the figure below:

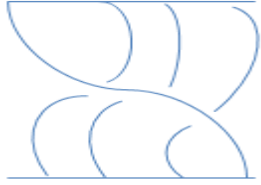

**Figure 3.** A representation of a general braid.

It can be immediately noticed that this picture is essentially the same as the figure 2, the remarkable difference is that two types of crossing, overcrossing and undercrossing, of the strands are distinctively highlighted. To relate the crossing of stocks diagram with the braid image, a simple convention designed to clear distinguishes between the two types of crossing in the case of the stocks time series of prices, has to be made. For every crossing of two neighbor stocks the difference between the price after and before the crossing are calculated for both stocks. The differences are taking in modulo since only the net amounts are considered, such that:

$$\Delta_{Stock\ i} = |P_{before\ crossing} - P_{after\ crossing}|, \qquad (1)$$

$$\Delta_{Stock\ i+1} = |P_{before\ crossing} - P_{after\ crossing}|, \qquad (2)$$

where P is the price of the stock. The stock with the higher difference it will come over and the stock with the smaller difference will be under in a stock crossing.

The two cases that can arise are:

- $\Delta_{Stock\ i} > \Delta_{Stock\ i+1}$ - the stock $i$ is crossing over the stock $i+1$, in which case the stocks crossing will be called to be an ***overcrossing of stocks***,
- $\Delta_{Stock\ i} < \Delta_{Stock\ i+1}$ - the stock $i$ is crossing under the stock $i+1$, in which case it will be called to be an ***undercrossing of stocks***.

The two situations in discussion for the crossing of two stocks are exemplified in the figure 4:

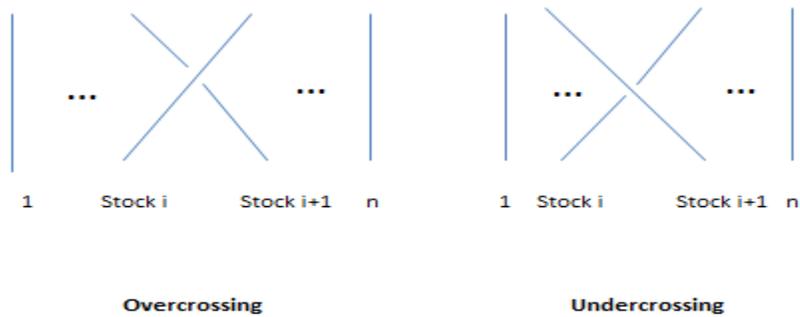

**Figure 4.** Overcrossing and undercrossing of stocks.

The explicit functionality of this schema can be shown in a simple example coming from the real market price quotations of stocks. Let's get back to the figure 1 b) and analyze the first crossing of stocks that show up at 05/21/2013 between stocks HD and WMT. Their Δ differences are:

$$\Delta_{HD}= |77{,}39 - 77{,}40| = |-0{,}01| = 0{,}01$$

$$\Delta_{WMT}= |78{,}71 - 76{,}76| = |1{,}95| = 1{,}95$$

such that $\Delta_{HD} < \Delta_{WMT}$ and it is an undercrossing between these two stocks.

Evaluating all the crossings of stocks in the diagram in figure 1b) the representation of the stock market (here only for 4 stock components) as a braid is shown in figure 5.

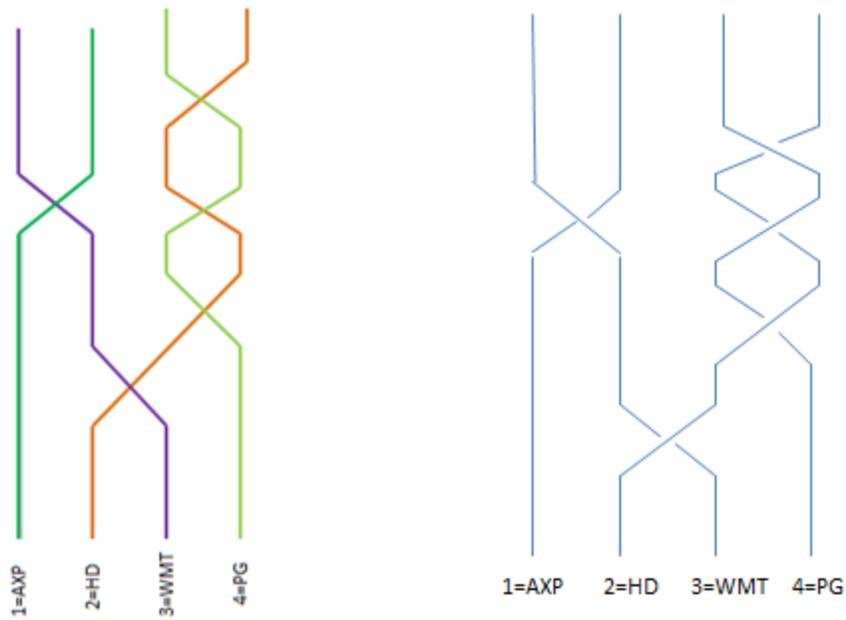

**Figure 5**. The crossings of stocks diagram and its braid diagram counterpart for 4 stocks components of DJIA index.

Mathematically braids form a group under concatenation and it remains true in the stock market circumstances. The concatenation of two braids is shown in figure 6.

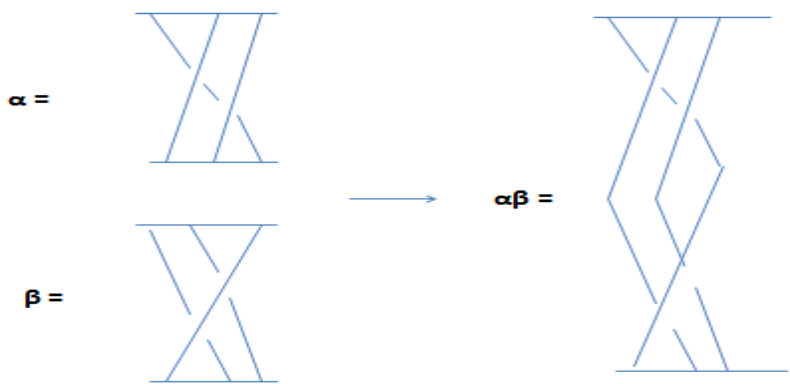

**Figure 6**. Concatenation of two braid.

Not entering in the algebra of braid group, the picture in figure 6 simply states that two braids can be added, which is of good sense when it comes to stock market.

4. **The stock market "whispers" – stock braids words**

It was stated in the latter sections that braids of stocks can be represented by n-strands that intertwining to form two types of crossing, overcrossing and undercrossing. The convention is to note the *overcrossing* with $\sigma_i$ in case the strand generated by the price quotations of stock $i$ passes over its right neighbor strand $i + 1$ and the *undercrossing* with $\sigma_i^{-1}$, in a reverse situation. The two types of crossings become the **generators of the stocks braid** and are sketched in figure 7.

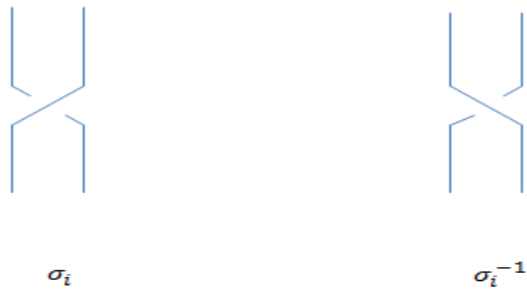

**Figure 7** . The braid generators.

The ordered succession in time of $\sigma_i$ and $\sigma_i^{-1}$ generators constitutes a **braid word**. To illustrate with an example the notion of braid word let's consider the generators for every crossing in figure 8. The braid word that results is:

$$W = \sigma_2\sigma_3\sigma_3\sigma_3^{-1}\sigma_1^{-1} = \sigma_2\sigma_3^2\sigma_3^{-1}\sigma_1^{-1} . \tag{3}$$

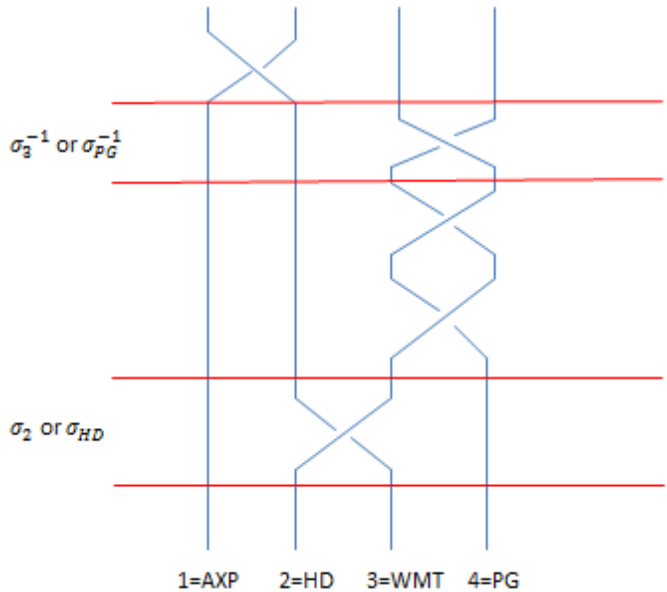

**Figure 8.** Braid generators in the notation adapted to stock market.

In the stock market activity the finance professionals are interested in moving of stock prices such that the notation of braid generators changes to account the needs in financial information.

Assuming the convention of stock notation for braid generators the braid word for the diagram in the figure 8 becomes:

$$W = \sigma_{HD}\sigma_{HD}\sigma_{PG}\sigma_{PG}^{-1}\sigma_{WMT}^{-1} = \sigma_{HD}^{2}\sigma_{PG}\sigma_{PG}^{-1}\sigma_{WMT}^{-1} \tag{4}$$

What the market "whispers" to investors by the word above? Actually a market analyst can extract valuable information from this "weird" braid word formulation. It immediately can be noticed that stock HD formed two overcrossings meaning that its price follows a bullish trend. PG after some crossings with HD finally find a bearish path and so the WMT stock. Taking a look at figure 1 b) just to compare these results with the real price quotations it could be seen that:

- The price of PG went up from 77,88 at 05/15/2013 to 79,08 at 06/03/2013;
- The price of PG went down from 80,68 to 77,66 at the same period of time;

- The price of WMT fall from 79,86 to 75,69 for the same period

which is in a perfect accord with what market "whisper" to investors in the simple word (4).

5. **Braids closure - knotting the stocks in the market**

According to Alexander Theorem every closed braid is a ***knot***. A closure of a braid can be obtained by simply gluing together its ends as is shown in figure 9.

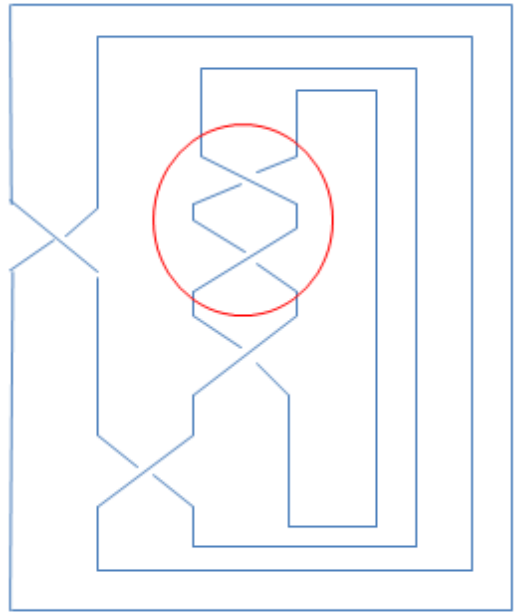

**Figure 9**. Closure of the stock market braid that reside in formation of a knot.

There are some methods to simplify the diagram of a knot that consist in removing the series of crossings that let the knot unchanged. The simplifying methods are called **Reidemeister moves** (see [11], [12]) and there are 3 types of such moves. In the approach of stocks knots only the Reidemeister move II will be used and the figure 10 shows a diagram of it.

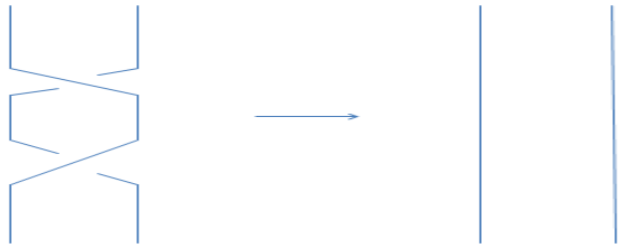

**Figure 10**. Reidemeister move II.

In the red circle in figure 11 it can be noticed that a Reidemeister move II is suited to simplify the knotted stocks, such that after applying the move the knot diagram can be depicted as:

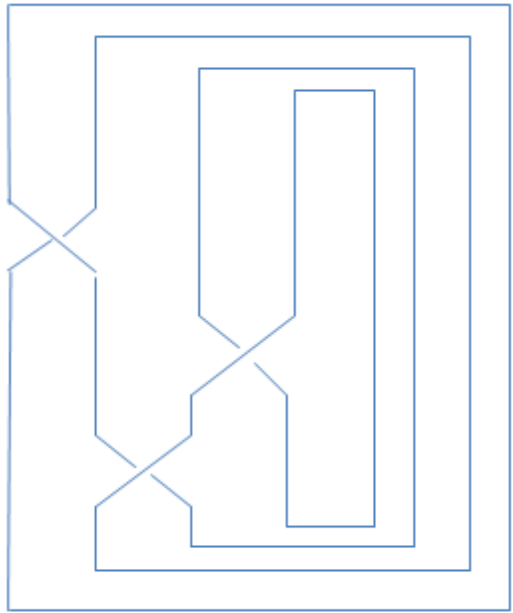

**Figure 11**. Simplified diagram for knotted stocks.

Having the stocks knotted is important to know what kind of knot stock market created in the process of price quotations. The next section will provide the algorithm to distinguish between diverse knots that stocks could create.

## 6. Distinguishing between knotted stocks – polynomial invariants

The attempts to ask the important question of distinguishing between knots find the first response in the work of J. Alexander. Alexander's method was heavily in use, but a surprisingly relation found by Conway made the knots theory a fashion in mathematics. The relation that Conway found is called the ***skein relation*** and is a polynomial that calculated it could segregate between types of knots. The knots theory flourish in the mid '80 when the most celebrated polynomial scheme was discovered by V. Jones. The applications ***of Jones Polynomials*** are vast and spread over numerous branches of research, from quantum physics to biology, and now finance.

There are many other, more complicated knot polynomials but in this paper the discussion is restricted at the two prior mentioned because there are simple and suffice to explaining the stock market behavior as a knot.

Prior to enter deeper in description and calculation of knot polynomials a small adjustment to the crossing of stocks interpretation should be make. Since knot polynomials refer to oriented knots an orientation should be given to the knotted stocks. The orientation is naturally that of the time flowing such that an ***oriented crossing of stocks*** is that sketched in the figure 12

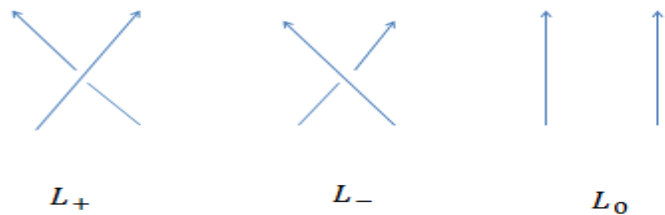

**Figure 12**. Types of oriented stocks crossing.

It could be notice from the figure above that, for convenience in exploring further the stock market knots the crossings were labeled $L_+$ for an overcrossing, $L_-$ for the undercrossing and $L_0$ when no crossing appear. This notation is the usual in knot theory being useful in defining the skein relation for the knot polynomials.

**The Alexander-Conway polynomial** $\nabla_L(z)$ of a knot K as it was stated earlier is based on the skein relation Conway discovered. The polynomial is obtained by applying the skein relation to every crossing of a knot until only unknots remain. This method of analyzing the type of a knot remains valid also in the case the Jones polynomial.

The skein relation for the Alexander – Conway polynomial is:

$$\nabla_{L_+}(z) = \nabla_{L_-}(z) + z\nabla_{L_0}(z), \tag{5}$$

where $z$ will be shifted to $t$ in the classical **Alexander polynomial** $\Delta_L(t)$ by the change: $z = \left(\sqrt{t} - \frac{1}{\sqrt{t}}\right)$ such that the skein relation will be :

$$\Delta_{L_+}(t) = \Delta_{L_-}(t) + \left(\sqrt{t} - \frac{1}{\sqrt{t}}\right)\Delta_{L_0}(t). \tag{6}$$

**The Jones polynomial** $V_L(t)$ of a knot K is obtained by following the same method as in the case of Alexander polynomial but this time using a skein relation of the type:

$$\frac{1}{t}V_{L_+}(t) = tV_{L_-}(t) + \left(\sqrt{t} - \frac{1}{\sqrt{t}}\right)V_{L_0}(t). \tag{7}$$

From this skein relation it is trivial cu calculate the two types of knot as:

$$V_{L_+}(t) = t^2 V_{L_-}(t) + t\left(\sqrt{t} - \frac{1}{\sqrt{t}}\right)V_{L_0}(t) \tag{8}$$

$$V_{L_-}(t) = t^{-2}V_{L_+}(t) + t^{-1}\left(\sqrt{t} - \frac{1}{\sqrt{t}}\right)V_{L_0}(t) \tag{9}$$

These skein relations are accompanied by an important axiom stated that in the case of the trivial knot (the unknot):

$$\Delta_L(t) = V_L(t) = V_O = 1 \tag{10}$$

For Jones polynomials I will add another result, that will help in calculation further some knots coming from the stock market configuration. The result, however easy to be calculated is:

$$V_{OO} = -\left(\sqrt{t} + \frac{1}{\sqrt{t}}\right) \tag{11}$$

and is the Jones polynomial for two unknotted.

Having said all these, for the sake of completeness and exemplification the polynomial for some knotted stocks is calculated. It may not be obvious but the stocks knot in figure 13 is nothing else than the trivial knot.

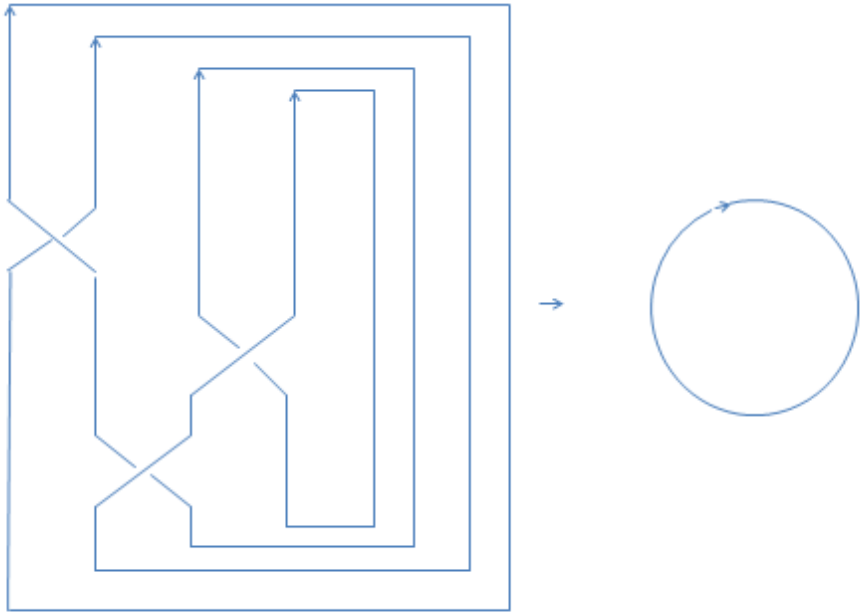

**Figure 13**. The unknotted stock market.

In this case the relation () applies and although it looks complicated, the stock market is unknotted.

Let's get back to the initial stock market configuration in figure 1 b) and choose this time the interval from 5/15/2013 to 6/7/2013 for the quotations of the stocks prices. After applying the Reidemeister move II for some crossings of stocks the final knot diagram is shown to the left of figure 14 along with a more intuitive picture representing the same knot shape to the right.

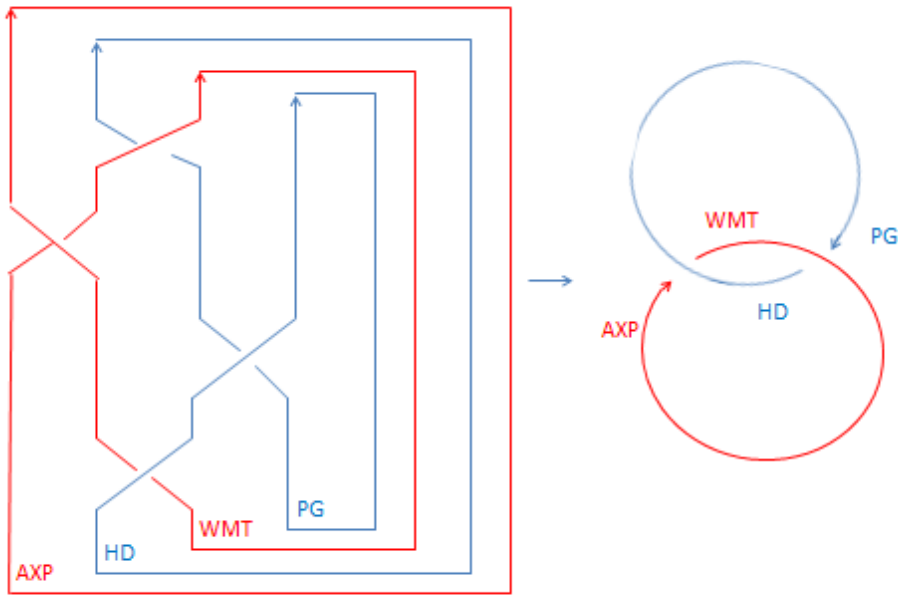

**Figure 14**. The stock market is linked.

The last crossing of the knot in figure 14 is decomposed according to the Jones skein relation

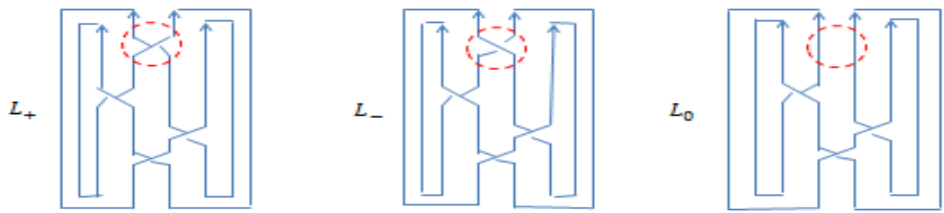

**Figure 15**. The representation of Jones skein relation.

Such that the Jones skein relation is:

$$t^{-1}V_{L_+} - tV_{L_-} = \left(\sqrt{t} - \frac{1}{\sqrt{t}}\right)V_{L_0} \tag{12}$$

From the image of the knots in figure 15 it should be said that $L_0$ is unknot as it was stated latter and $L_-$ is formed by two unknots, one in top of the other, such that $L_- = L_{OO}$. The skein relation (12) becomes:

$$t^{-1}V_{L_+} - t\left(-\sqrt{t} - \frac{1}{\sqrt{t}}\right) = \left(\sqrt{t} - \frac{1}{\sqrt{t}}\right) \tag{13}$$

After some simple calculation the Jones polynomial for the stock market (in the chosen formation of 4 stocks) is:

$$V_{L_+} = -t^{5/2} - t^{1/2} \tag{14}$$

The representation of the stock market in this case is the **positive Hopf link**.

The Alexander polynomial will slightly differ from the Jones polynomials since in for this knot invariant $\nabla_{OO} = \Delta_{OO} = 0$. The scheme for calculation is shown in the figure 15 and is the same like in the Jones polynomial exemplification.

The calculation according to the Alexander-Conway skein relation is as follow:

$$\nabla_{L_+} = 1\,\nabla_{OO} + z\,1\,\nabla_O$$

$$= 0 + z = z \tag{15}$$

and in classical Alexander notation:

$$\Delta_{L_+} = t^{1/2} - t^{-1/2} \tag{16}$$

This result is trivial in the mathematical literature, still for knots having more crossings there are tables containing polynomials at which a polynomial that result in calculating the market stocks crossings can be compared with. In such way a polynomial can be assimilated with a certain knot. The appendix illustrates a fragment of a knots table taken from the **Encyclopedia of Mathematics** site.

7. **Relation with statistical mechanics and quantum computing**

Now having the stock market all knotted and the resulted knot find itself classified in tables, the question is what good such a representation is for. Leaving aside the beauty of the idea of a **topological stock market,** it remains the eternal Wall Street "show me the money!" aspect. The question resumes to "how can I profit from this knots, braids or geometric shapes?"

I would like to stretch out from the beginning that the mathematical aspects in this paper are part of a larger research project looking to apply topology and geometry to financial realities and much work is still to be done in this particular field. As a result large parts remained yet unexplored. Still, some speculations about possible applications of stocks topology can be made.

I will emphasize here the remarks of V. Jones about the connection of the Jones polynomials to statistical mechanics. Although, not explored yet as I stated earlier, it looks to be a promising candidate in forecasting mini-flash crashes and to some extent financial crisis.

Many financial analysts came to the conclusion that flash-crashes and by extension financial crisis are the market phase transitions from a smooth behavior to a regime marked by discontinuities in prices movements.

I will quote here a short section of an article Mark Buchanan wrote on the personal blog:

*"…a key determinant of market dynamics is the diversity of participants' strategic behavior. Markets works fairly smoothly if participants act using many diverse strategies, but break down if many traders chase few opportunities and use similar strategies to do so. Strategic crowding of this kind can cause an **abrupt phase transition** from smooth behavior into a regime prone to sharp, virtually discontinuous price movements. One fairly recent study suggested that **high-frequency trading** may be pushing modern markets through such a phase transition, with the breakdown of the continuity of prices movements (lots of mini-flash crashes) being one major consequence. The underlying phase transition phenomenon may therefore be quite relevant to policy. I know of nothing in traditional equilibrium economic analysis that describes this kind of phase transition*."

Turning to Jones polynomial, as V. Jones puts it in [4], [10], referring to phase transitions models (Potts and Ice-type) in statistical mechanics:

*"thus the Jones polynomial of a closed braid is the partition function for a statistical mechanical model"* and

*" It is a miracle that the choice…gives the Jones polynomial of the link defined by D as its partition function "*.

Knowing the Jones polynomial for knotted stocks in the market could directly define the partition function of a Potts model associated to stock market, such that it could anticipate the mini-flash crashes related to market phase transitions generated by high-frequency trading activities. It also at a larger scale, anticipates the financial markets phase transition to a crisis like the one experience in 2007-2008.

Anticipating the stock market phase transitions is equally important to market participants and also to market regulators that could create policies to prevent financial crisis.

I would not end this section without mentioning an intriguing resemblance of the braided stock market with the newly discovered ***Topological quantum computer***. As the topology of the stock market is constructed from braiding the stocks, the quantum circuits in the topological quantum computer are constructed from braiding of anyons (see [5], [8], [9]), such that their invariant polynomials (the Jones polynomial) are the qubits. This association of facts led to the astonishing conclusion that the stock market stats could be quantum encoded by qubits resulted from braiding the stocks, such that the stock market itself is a topological quantum computer. Leaving aside these simplest speculations, it should be said that in the dawn of cryptographic money era, opened by bitcoin, this result might have more important connotations then the speculations above shows.

These issues remain open for now and it will be explored in future research.

8. **Conclusions**

The paper [1] proposes a new method of interpreted the behavior of market indexes or a particular choice of a portfolio of stocks. In the above mentioned paper the stocks composing the DJIA index are arranged in a table in ascending order of price quotations from the right to the left. To keep track with the original time series of any stock every stock price quotation time series is colored with a different color.

Under this scenario of arranging the stocks prices, a beautiful diagram showing the crossings between neighbor stocks can be depicted. This particular vision of stocks unlocks some remarkable mathematical objects, such as permutations, matroids, braids and knots.

The present paper introduces a topological approach of the stock market intermediated by braids and knots that stocks form in the process of price quotation. The crossings of stocks are categorized as overcrossings and undercrossings, and form the generators for the building the stocks braid.

The braid generators in their time succession "write words" that could give investors important insights about stock market state.

Gluing together the ends of a stocks braid it lead to the formation of a knot, a beautiful topological structure that became a mathematical fashion at the end of the last century once V. Jones discovered its polynomial invariant ( see[2]). The Jones polynomial and Alexander-Conway polynomial are used here to distinguish between knotted stocks. Knots are classified by their polynomial in tables and a fragment of such table is shown in annex.

The topological aspects stock market could appear in many financial applications but only two of them are briefly sketched in the present paper. Taken into account a remark of V. Jones that Jones polynomial of a close braid is the partition function of a statistical mechanical phase transition model the polynomial of knotted stocks could provide a clear way to anticipate the transition of financial markets to the phase that leads to crisis.

The resemblance of the braiding stocks with the logic gates of the topological quantum computer is a second hint about the ways braided and knotted stocks topology could be applied to financial realities. Under this scenario the Jones polynomial of some stocks knot would represent qubits that encode the stock market quantum states in the way that only topological quantum computers could decode. It might be a simple speculation, but in the down of the cryptographic money era, opened by bitcon, it could have some important connotations.

**Appendix.**

Classification of knots by their Alexander polynomial – a fragment of the knots table.